\documentclass[conference]{IEEEtran}
\IEEEoverridecommandlockouts
\usepackage{cite}
\usepackage{amsmath,amssymb,amsfonts}
\usepackage{algorithmic}
\usepackage{graphicx}
\usepackage{textcomp}
\usepackage{xcolor}
\usepackage{balance}
\def\BibTeX{{\rm B\kern-.05em{\sc i\kern-.025em b}\kern-.08em
    T\kern-.1667em\lower.7ex\hbox{E}\kern-.125emX}}
\usepackage{enumitem}
\usepackage{tikz}
\usetikzlibrary{shapes.geometric, positioning}

\tikzstyle{block} = [rectangle, minimum width=2.5cm, minimum height=1cm, text centered, draw=black, font=\large]
\tikzstyle{tallblock} = [rectangle, minimum width=2.5cm, minimum height=2.5cm, text centered, draw=black, font=\large]
\tikzstyle{circleblock} = [circle, draw, minimum size=1.5cm, text centered, font=\large]
\tikzstyle{leftarrowblock} = [block, append after command={
    \pgfextra \draw[black, fill=black]
        (\tikzlastnode.west) ++(-0.4, 0) -- ++(0.4, 0.5) -- ++(0, -1) -- cycle;
    \endpgfextra
}]
\tikzstyle{tallleftarrowblock} = [tallblock, append after command={
    \pgfextra \draw[black, fill=black]
        (\tikzlastnode.west) ++(-0.4, 0) -- ++(0.4, 1.25) -- ++(0, -2.5) -- cycle;
    \endpgfextra
}]
\tikzstyle{rightarrowblock} = [block, append after command={
    \pgfextra \draw[black, fill=black]
        (\tikzlastnode.east) ++(0.4, 0) -- ++(-0.4, 0.5) -- ++(0, -1) -- cycle;
    \endpgfextra
}]

\begin{document}
\begin{sloppy}

\title{Generative AI and Empirical Software Engineering:\\A Paradigm Shift}

\author{\IEEEauthorblockN{Christoph Treude}
\IEEEauthorblockA{\textit{School of Computing and Information Systems} \\
\textit{Singapore Management University}\\
Singapore, Singapore \\
ctreude@smu.edu.sg}
\and
\IEEEauthorblockN{Margaret-Anne Storey}
\IEEEauthorblockA{\textit{Department of Computer Science} \\
\textit{University of Victoria}\\
Victoria, Canada \\
mstorey@uvic.ca}
}

\maketitle

\begin{abstract}
The adoption of large language models (LLMs) and autonomous agents in software engineering marks an enduring paradigm shift. These systems create new opportunities for tool design, workflow orchestration, and empirical observation, while fundamentally reshaping the roles of developers and the artifacts they produce. Although traditional empirical methods remain central to software engineering research, the rapid evolution of AI introduces new data modalities, alters causal assumptions, and challenges foundational constructs such as ``developer'', ``artifact'', and ``interaction''. As humans and AI agents increasingly co-create, the boundaries between social and technical actors blur, and the reproducibility of findings becomes contingent on model updates and prompt contexts. This vision paper examines how the integration of LLMs into software engineering disrupts established research paradigms. We discuss how it transforms the phenomena we study, the methods and theories we rely on, the data we analyze, and the threats to validity that arise in dynamic AI-mediated environments. Our aim is to help the empirical software engineering community adapt its questions, instruments, and validation standards to a future in which AI systems are not merely tools, but active collaborators shaping software engineering and its study.
\end{abstract}


\section{Introduction}

The software engineering research and industry communities are undergoing a profound transformation driven by the rapid development and adoption of large language models (LLMs) and agentic AI technologies. Many consider generative AI to be the most disruptive innovation in software engineering since the advent of the Internet~\cite{playbook}, with the potential to fundamentally change the way software is developed, evolved, and maintained~\cite{rao2025software}. Even conservative perspectives recognize that this shift has already brought substantive changes to development practices~\cite{notmagic}, influencing delivery speed, software quality, collaboration patterns, and the overall developer experience.

Beyond changes to practices, LLM-driven tools and agents are reshaping the roles of developers, users, and researchers. The boundaries between these actors are becoming increasingly blurred, enabling new approaches to designing, building, and studying software systems. Developers now collaborate with AI systems such as GitHub Copilot, ChatGPT, and Claude Code; organizations experiment with autonomous developer agents; and repositories introduce orchestration files such as \texttt{.agents.md} and \texttt{.copilot-instructions} that govern the behavior of AI teammates~\cite{mohsenimofidi2025context}. These developments challenge the empirical methods traditionally used in software engineering research. Moreover, findings that were valid at one point in time may quickly become outdated as models evolve, a challenge sometimes referred to as evaluation drift~\cite{rao2025software}.

Not only are the methods being disrupted, but so are the research questions we ask. The influence of technological change on research is a long-standing concern in fields such as media studies and digital anthropology~\cite{digitalanthro}. Marshall McLuhan~\cite{mcluhan1977laws} famously proposed four laws for analyzing the effects of new technologies, and Storey et al.~\cite{playbook} have applied these to the case of generative AI in software engineering:

\begin{itemize}
\item \textbf{What does generative AI enhance or amplify?} It amplifies many software engineering tasks~\cite{hou2023large}, such as automating the writing of low-level code that previously required substantial human effort, and increasingly orchestrates larger tasks such as generating test suites, documentation, and even pull requests.
\item \textbf{What does the technology make obsolete?} Platforms such as Stack Overflow are seeing reduced usage as developers turn to AI for instant assistance~\cite{xu2023we}, reducing the dependency on community-driven resources and reshaping knowledge-sharing ecosystems.
\item \textbf{What does the technology retrieve that had been obsolesced earlier?} Generative AI brings back conversational and chat-based interfaces for technical assistance~\cite{shawar2007chatbots}, seamlessly integrating them into development workflows where textual dialogue once played a secondary role.
\item \textbf{What does the technology reverse or flip into when pushed to extremes?} With increasing reliance on AI, foundational skills such as learning programming from first principles and understanding the underlying abstractions risk being neglected~\cite{hicks2024new}. Over-trust in AI suggestions may also reverse productivity gains by introducing subtle errors and reducing developer confidence.
\end{itemize}

These four laws offer a useful lens to understand the disruptions and opportunities caused by the adoption of LLMs. Yet, they also highlight a deeper challenge: As our tools change, so must our research practices. The nature of our data is shifting, the definition of software artifacts is expanding, and the boundaries of human and machine agency are being redrawn. Empirical research must therefore grapple with new kinds of research phenomena, methodological adaptations, theoretical reframings, and emerging threats to validity.

This vision paper explores how LLMs and agentic AI are reshaping empirical software engineering. Our contribution differs from prior perspectives such as the Copenhagen Manifesto~\cite{russo2024generative} and the FM+SE Vision 2030 workshops. While those initiatives primarily target practitioners and community leaders, emphasizing responsible adoption and ethical guidelines for generative AI in software engineering, our vision specifically addresses the empirical research community. We focus on how constructs, methods, data, and validity criteria in empirical software engineering must evolve to rigorously study AI-mediated software practice. Specifically, we examine: (1) the evolving phenomena and research questions that demand attention; (2) the need to adapt research methods and theories to address the unique characteristics of AI-driven systems; (3) the emergence of new data types and the opportunities and limitations they present; and (4) the threats to validity stemming from the non-deterministic, dynamic nature of AI tools and their integration into development workflows.

Through this examination, we aim to prepare the empirical software engineering community for a future in which AI is not merely a tool but an active actor~\cite{latour2007reassembling} in the software development process: participating in analysis, design, implementation, and research itself. As adoption accelerates~\cite{ferino2025novice} and autonomous agents become integral to development ecosystems, addressing these challenges becomes increasingly urgent.

\section{The Research Phenomena We Study and the Questions We Ask}

The adoption of LLMs and agentic AI systems is fundamentally transforming the landscape of software engineering research. Mustafa Suleyman, in his book ``The Coming Wave'', describes generative AI as a ``general-purpose technology''~\cite{mustafa}, on par with foundational innovations such as fire, the wheel, and the computer. Such technologies have historically triggered ``Cambrian explosions'' of innovation, reshaping how humans live, work, and interact. Similarly, when combined with other emerging technologies such as synthetic biology, robotics, and augmented reality, generative AI has the potential to drive unprecedented transformations across technical and social domains.

Latour's actor-network theory~\cite{latour2007reassembling} reminds us that innovation often emerges within unstable categories and uncertain boundaries. Researchers must ``follow the actors'' to understand how collective existence is redefined. As LLMs become embedded in the everyday practices of software engineering, long-standing constructs such as ``developer'', ``source code'', and ``artifact'' are becoming more fluid, prompting a reexamination of what, who, and how we study.

These shifts are particularly evident in the evolving relationships among developers, tools, and end users. Developers increasingly rely on AI to generate, refine, and explain code, shifting their efforts from writing to orchestrating and evaluating solutions~\cite{xiao2025self, butler2025dear}, as seen in emerging AI-native IDEs such as Windsurf, which combine conversational prompting, autonomous agents, and code-context awareness within the development environment~\cite{terragni2025future}. End users, in turn, are beginning to perform development-like tasks with the help of conversational interfaces, and emerging practices such as ``vibe coding''~\cite{pimenova2025good} -- specifying high-level intent or style rather than concrete requirements -- further blur the line between ``user'' and ``developer''. Meanwhile, generative AI tools and autonomous agents now directly contribute to the creation of software artifacts, such as commit messages, bug reports, design diagrams, or documentation, challenging the conventional division between tools and collaborators. For example, if GitHub Copilot is ``your AI pair programmer'', as its tagline suggests, then is it the driver or the navigator in this new form of pair programming~\cite{treude2025bot}? The question becomes even more complex when agents review or supervise one another’s work, as seen in emerging multi-agent software ecosystems~\cite{he2025llm}.

Naur’s view~\cite{Naur} that software consists of theories in the developer's mind, rather than just in code, becomes especially relevant in the age of prompt-based and agent-assisted development. Developers now externalize parts of their mental models into prompt instructions or orchestration files, delegating reasoning tasks to AI collaborators~\cite{treude2025interacting}. New interaction modalities, including natural language prompts, sketches~\cite{programmingbyexample}, and gestures in immersive environments~\cite{VRsoftware}, further stretch the traditional definitions of ``coding'', ``testing'' and ``software''. Increasingly, programming involves shaping a dialogue with an intelligent system rather than manipulating syntax alone.

As Latour suggests, when innovation accelerates, we must revisit the constructs that define our field and observe how new actors emerge and interact. For empirical software engineering researchers, this involves studying phenomena such as the integration of AI into the inner development loop, the shifting balance of control and trust between humans and agents, the reconfiguration of team dynamics, and the emergence of hybrid workflows combining human intent with AI-generated context. Moreover, the increasing use of synthetic data and AI-simulated participants introduces new opportunities for experimentation, but also raises concerns about ecological validity and authenticity.

These transformations also have broader ethical and social implications. Generative AI systems learn from interaction traces that may unintentionally encode biases or amplify inequities present in historical data. Researchers must remain vigilant about these effects and develop methods to audit, mitigate, and contextualize them. Issues of authorship, accountability, and transparency -- long studied in software engineering -- now resurface in new forms as AI begins to participate in both software creation and research processes.

In short, the adoption of LLMs and agentic AI is not merely an extension of existing software practices; it redefines the very phenomena that empirical software engineering seeks to understand. By posing new research questions and updating our conceptual frameworks, we can ensure that empirical software engineering research remains relevant, rigorous, and forward-looking in this new era.

To guide future work in this evolving context, we recommend:

\begin{itemize}
\item Continuously refining definitions of constructs, such as ``developer'', ``artifact'', and ``coding'', to reflect their evolving roles in AI-mediated environments and to capture interactions involving both human and non-human actors.
\item Focusing on questions that explore the unique dynamics of human-AI collaboration, including effects on creativity, trust, collaboration, explainability, and knowledge transfer across roles and time.
\item Conducting longitudinal and adaptive studies to examine how AI adoption influences practices, team structures, and skill development over time, accounting for changes in model versions and capabilities.
\item Collaborating across disciplines, including cognitive science, education, sociology, and HCI, and engaging with diverse developer communities to surface novel insights into the socio-technical impacts of generative AI.
\end{itemize}

\section{Research Methods and Theories}

The rise of LLMs and agentic AI challenges the traditional methods and theories that have long supported empirical software engineering research. To remain relevant and rigorous, these approaches must evolve to address the new socio-technical dynamics introduced by AI systems, the unprecedented pace of their evolution, and the expanding scale and heterogeneity of the data they produce.

Historically, empirical software engineering has relied on a combination of quantitative, qualitative, and mixed-method approaches to analyze software artifacts and practices. Although these foundations remain vital, the integration of LLMs introduces new methodological complications and epistemic risks:

\begin{itemize}
\item \textbf{Quantitative methods:} The non-deterministic nature of AI outputs complicates statistical analysis, causal inference, and reproducibility. Researchers must account for both output variability and the rapid evolution of underlying models, which can cause evaluation drift between data collection and publication. Transparent reporting of model versions, parameters, and prompts is therefore essential to maintain comparability across studies~\cite{baltes2025guidelines}.
\item \textbf{Qualitative methods:} As the boundaries between human and AI agency blur, it becomes harder to isolate human intent, cognition, and behavior. Traditional observational or interview-based techniques to study developer–system interactions must be adapted to treat AI not as a background tool but as an active participant. This calls for protocols that capture both human reflections and AI reasoning traces, as well as reflexive analysis of researcher-AI interactions when AI assists in data interpretation. Moreover, as multi-agent workflows become more common, qualitative inquiry must also account for AI-AI interactions -- how autonomous agents critique, coordinate with, or override one another -- since these exchanges increasingly shape the emergent dynamics of software development.
\item \textbf{Mixed-method designs:} Embedded mixed-method approaches~\cite{poth} remain well suited to capture the interplay between human and AI actors in software workflows. Yet researchers must now manage evolving AI outputs, hybrid artifacts that combine human and machine contributions, and the risk that automated processes may obscure important socio-technical signals. Triangulation between behavioral data, conversational logs, and post-hoc interviews becomes key to contextualizing results.
\end{itemize}

These changes require a larger transformation in study design. AI-mediated environments are dynamic and continuously updated; findings valid today may lose relevance as models or datasets shift. Longitudinal, adaptive, and replication-oriented designs will become increasingly necessary to sustain validity. The notion of “frozen datasets” must give way to “living studies” that document temporal variation in model behavior.

The complexity of the impact of AI also requires deep interdisciplinary collaboration. Insights from other domains can enrich empirical research in several complementary ways. For example, education researchers can examine how AI influences the learning and development of programming skills, a known determinant of developer success~\cite{hicks}. Cognitive scientists can explore how AI alters mental models, attention patterns, and problem-solving strategies, in ways comparable to how the Internet reshaped human cognition~\cite{shallows}. Social scientists can investigate how AI reconfigures team dynamics, organizational hierarchies, and systemic equity, while economists can assess productivity trade-offs, cost structures, and emergent markets for AI-generated code.

This interdisciplinary lens also supports the evolution of theoretical frameworks. Existing theories, such as McLuhan's media laws~\cite{mcluhan1977laws}, Latour's actor network theory~\cite{latour2007reassembling}, and the Storey et al. playbook~\cite{playbook}, provide valuable perspectives for analyzing the evolving relationships among developers, tools, and artifacts in the AI era. However, these theories must be extended or hybridized to capture the role of AI as an autonomous collaborator that can generate, evaluate, and even negotiate meaning. Emerging constructs such as construction-trace analysis, dual-loop active learning, and human-AI co-work frameworks offer promising foundations for theorizing about feedback cycles where humans supervise, correct, and retrain AI systems in context.

Generative AI is not only a research subject: it is rapidly becoming a research instrument. Recent studies demonstrate that AI can assist in coding large volumes of qualitative data, such as analyzing user feedback or identifying themes in developer-AI interactions~\cite{ahmed2025can, bano2024large}. LLMs can also support mixed-method analysis at scale, simulate participant behavior in controlled experiments~\cite{steinmacher}, and summarize large corpora of development traces to accelerate theory building. However, these capabilities introduce new dependencies: models may embed unknown biases, training data provenance may be opaque, and fine-tuned behavior can shift without notice.

Ethical and epistemic concerns further complicate AI-assisted research. AI systems may reproduce societal biases, generate misleading explanations, or fabricate evidence. Over-reliance on AI outputs risks displacing critical human judgment, particularly when AI automates interpretive tasks that traditionally required reflexivity and domain expertise. Responsible integration of AI into research processes thus requires transparency, reproducibility, and human oversight at every interpretive stage.

As the distinction between technical systems and social processes continues to blur, empirical researchers need frameworks that explicitly account for AI as a socio-technical actor. This includes understanding how AI-generated data shapes our interpretations of human behavior, how AI mediates cognitive and collaborative processes within software teams, and how the researcher-tool relationship itself evolves when AI participates in knowledge construction.

To address these challenges, we recommend:

\begin{itemize}
\item Developing new metrics, benchmarks, and reporting standards to evaluate AI-mediated workflows, artifacts, and research reproducibility.
\item Collaborating with researchers from other disciplines to refine empirical methods, theoretical perspectives, and validation strategies for hybrid human-AI contexts.
\item Establishing transparent practices and safeguards to surface biases, disclose model configurations, and clarify how AI is applied in each research stage.
\item Embracing iterative, adaptive, and longitudinal study designs that evolve with the rapid pace of AI innovation and enable replication across shifting models, prompts, and contexts.
\end{itemize}

\section{The Data We Study}

LLMs and agentic AI systems are reshaping not only software engineering practices and tools but also the very nature of the data available for empirical research. The importance of studying software engineering data has long been recognized. In the 1996 editorial for the inaugural issue of the Empirical Software Engineering journal, Harrison and Basili defined the field as “the study of software-related artifacts for the purpose of characterization, understanding, evaluation, prediction, control, management, or improvement through qualitative or quantitative analysis”~\cite{harrison1996editorial}. Almost three decades later, this definition remains foundational. For example, Abou Khalil and Zacchiroli~\cite{abou2022software} conducted a meta-analysis of artifact mining in empirical software engineering, cataloging more than 3,000 studies that mined data such as bug reports, code reviews, commit metadata, forum posts, email data, source code, test data, and UML diagrams. They also identified common artifact combinations used for goals such as bug prediction and source-code classification.

However, in the era of LLMs, many of these artifacts are no longer produced solely by humans. Increasingly, they are generated or co-generated by AI tools. For example, AI systems now contribute to the creation of bug reports~\cite{shi2022buglistener}, code-review comments~\cite{lu2023llama}, commit messages~\cite{zhang2024automatic}, forum responses~\cite{zhong2024can}, and emails~\cite{cambiaso2023scamming}. They also generate source code~\cite{liu2024your}, test cases~\cite{xue2024domain}, and UML diagrams~\cite{conrardy2024image}. These artifacts introduce a new category of empirical data: AI-mediated artifacts, whose provenance, authorship, and intent are jointly distributed between humans and models.

These developments raise a critical question for the empirical software engineering community: in a landscape where AI generates a lot of data, what role should researchers play? Are we merely studying AI outputs, or can we understand and shape the socio-technical mechanisms that produce them?

To answer this, it is important to unpack the mechanics of generative systems. At the heart of AI models are deep neural architectures trained on large corpora of text from books, articles, websites, and code repositories. These models are activated through prompts, natural language queries or instructions, after which they generate outputs conditioned on their training data. This process yields a wide range of software artifacts, from code snippets and bug reports to commit messages, pull-request discussions, and design diagrams.

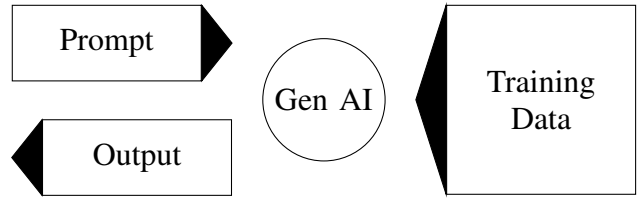
\begin{figure}[t]
\centering
\begin{tikzpicture}
\node (prompt) [rightarrowblock, yshift=-.5cm] at (0, 1) {Prompt};
\node (output) [leftarrowblock, below=of prompt, yshift=.5cm, xshift=.4cm] {Output};
\node (llm) [circleblock, right=of prompt, yshift=-0.75cm, xshift=-.2cm] {Gen AI};
\node (data) [tallleftarrowblock, right=of llm, xshift=-.2cm] {\parbox{2cm}{\centering Training \ Data}};
\end{tikzpicture}
\caption{New types of data amenable to empirical software engineering research in the era of generative-AI adoption: training data, prompts, and output.}
\label{fig}
\end{figure}

As shown in Fig.~\ref{fig}, empirical researchers now engage with three interconnected data sources: training data, prompts and interactions, and generative outputs. Together, these offer new opportunities for studying AI-mediated development.

\textbf{Training data.}
The composition and quality of the training corpora determine how models reason and respond. For example, Lin et al.~\cite{10.1145/3762183} showed that experience-aware oversampling improves informativeness and correctness in generated reviews without adding new data. Similarly, Nguyen et al.~\cite{nguyen2024encoding} demonstrated that incorporating version history data enhances the detection of code clones. In contrast, poorly curated data can establish bias. Treude and Hata~\cite{treude2023she} uncovered gender bias in models trained in software engineering text, while Sami et al.~\cite{sami2023case} found similar problems in image generation systems such as DALL-E-2. These findings highlight the need for systematic dataset auditing, documentation, and representativeness checks. Provenance analysis, which tracks where the data originates and how it influences downstream behavior, has become an essential methodological step.

\textbf{Prompts and interactions.}
The study of prompts has rapidly become a central theme in empirical software engineering. Building on previous work on how developers interact with bug reports~\cite{davies2014s}, commits~\cite{alali2008s}, and code reviews~\cite{li2017they}, researchers now examine how developers craft, refine, and negotiate prompts with AI systems~\cite{treude2025how}. Datasets such as DevGPT~\cite{xiao2024devgpt}, containing thousands of developer-ChatGPT conversations, reveal how prompt engineering evolves over time and how users calibrate trust through feedback cycles. Beyond static logs, future research should incorporate observational, longitudinal, and survey-based approaches to contextualize prompt behavior in real development settings.

\textbf{Generative outputs.}
AI-generated artifacts raise questions about correctness, completeness, and usefulness. Previous work has largely emphasized functional accuracy~\cite{liu2024no, yetistiren2022assessing}, but broader epistemic concerns are emerging. Storey et al.~\cite{playbook} and McLuhan’s media laws~\cite{mcluhan1977laws} remind us to consider what is lost when human-centric knowledge systems are replaced by AI mediation. Just as the rise of Stack Overflow improved productivity but reduced documentation diversity~\cite{parnin2012crowd}, the dominance of generative models may homogenize code, compress stylistic variation, and erode mentoring and collaborative learning opportunities.

Emerging data types such as configuration artifacts (\texttt{.agents.md}), execution traces of autonomous and multi-agent systems, and revision logs of AI-generated content create further opportunities for empirical analysis. They also pose challenges: data contamination between training and evaluation, privacy risks in conversational logs, and the need for sustained version tracking to ensure comparability across model generations.

To navigate these challenges and opportunities, we recommend lines of inquiry such as the following:

\begin{itemize}
\item Investigate which aspects of human coding practice may be overlooked or homogenized in AI-generated or AI-modified code.
\item Analyze which contextual cues are lost when bug reports or reviews are generated or pre-filtered by AI.
\item Examine how mentorship, authorship, and knowledge transfer change in AI-driven code-review workflows.
\item Compare the clarity, rationale, and completeness of AI-generated commit messages with human-written ones.
\item Evaluate how well AI-generated test oracles and summaries capture human judgment and intuition.
\item Study how community participation and learning dynamics shift as AI mediates or replaces traditional collaborative platforms.
\end{itemize}

\section{Reconsidering Threats to Validity}

The integration of LLMs and agentic AI into software engineering introduces new threats to validity that challenge established norms in empirical research. These threats arise from the dynamic and non-deterministic behavior of AI systems, the evolving nature of core software engineering constructs, and the increasing reliance on AI-generated data and tools within the research process itself.

\textbf{Construct validity} refers to whether a study accurately measures the concepts it intends to investigate. LLMs complicate this by shifting the meaning of foundational constructs. Terms such as ``developer'', ``source code'', and ``artifact'' evolve as AI systems take more active roles in workflows. For example, when natural language prompts begin to replace programming languages, the definition of ``coding'' itself becomes less clear. As AI blurs the line between social and technical actors, existing constructs may no longer capture the complexity of the phenomena being studied.

\textbf{Internal validity}, or the ability to establish causal relationships, is also affected. AI-generated outputs are inherently variable; even identical prompts can yield different responses across runs or model versions. This variability complicates replication and weakens causal inference. Frequent model updates lead to evaluation drift, where results obtained today may no longer hold tomorrow, compromising longitudinal stability and reproducibility.

\textbf{External validity} concerns the generalizability of study findings. Generative systems introduce several constraints in this regard. Studies relying on proprietary models may not generalize to other systems with different architectures or training data. Controlled settings may fail to capture the diversity of real-world AI interactions, particularly as tools spread across different developer communities. As LLMs continue to evolve rapidly, the need for ecologically valid research is becoming increasingly urgent.

LLMs also pose methodological and interpretive risks when used directly in research. Employing AI to code qualitative data or simulate participants may introduce hidden model assumptions. Since many systems are opaque, it can be difficult to trace how outputs were produced. This opacity affects interpretability and reproducibility. Moreover, over-reliance on AI tools may cause researchers to overlook subtle contextual cues that manual analysis would reveal. Therefore, transparency in model versions, prompts, and evaluation parameters becomes central to research validity.

To protect rigor in AI-related studies, we recommend:

\begin{itemize}
\item Prefer open or reproducible models to enhance transparency and facilitate replication.
\item Employ adaptive research designs that can accommodate shifting model behaviors and evolving tool capabilities.
\item Regularly revisit constructs to reflect changes in the software engineering landscape caused by AI.
\item Leverage mixed methods~\cite{poth, mixedthreats} to capture the technical and social dimensions of AI-mediated workflows.
\item Explicitly evaluate and report on the assumptions and biases present in AI systems used in each study.
\end{itemize}

\section{Concluding Remarks}

LLMs are transforming developers’ roles, reshaping artifacts, and redefining empirical research. As boundaries blur between human and machine, between artifact and interaction, and between subject and method, the empirical software engineering community must reassess its assumptions, adapt its practices, and embrace new lines of inquiry.  

\section*{Acknowledgements}

Many of the ideas presented in this paper were inspired by discussions at the Copenhagen Symposium on Human-Centered AI in Software Engineering.

\end{sloppy}
\balance

\begin{thebibliography}{10}
\providecommand{\url}[1]{#1}
\csname url@samestyle\endcsname
\providecommand{\newblock}{\relax}
\providecommand{\bibinfo}[2]{#2}
\providecommand{\BIBentrySTDinterwordspacing}{\spaceskip=0pt\relax}
\providecommand{\BIBentryALTinterwordstretchfactor}{4}
\providecommand{\BIBentryALTinterwordspacing}{\spaceskip=\fontdimen2\font plus
\BIBentryALTinterwordstretchfactor\fontdimen3\font minus \fontdimen4\font\relax}
\providecommand{\BIBforeignlanguage}[2]{{%
\expandafter\ifx\csname l@#1\endcsname\relax
\typeout{** WARNING: IEEEtran.bst: No hyphenation pattern has been}%
\typeout{** loaded for the language `#1'. Using the pattern for}%
\typeout{** the default language instead.}%
\else
\language=\csname l@#1\endcsname
\fi
#2}}
\providecommand{\BIBdecl}{\relax}
\BIBdecl

\bibitem{playbook}
M.-A. Storey, D.~Russo, N.~Novielli, T.~Kobayashi, and D.~Wang, ``A disruptive research playbook for studying disruptive innovations,'' \emph{ACM Transactions on Software Engineering and Methodology}, vol.~33, no.~8, pp. 1--29, 2024.

\bibitem{rao2025software}
H.~Rao, Y.~Zhao, X.~Hou, S.~Wang, and H.~Wang, ``Software engineering for large language models: Research status, challenges and the road ahead,'' \emph{arXiv preprint arXiv:2506.23762}, 2025.

\bibitem{notmagic}
T.~Leaver and S.~Srdarov, ``{ChatGPT} isn't magic: The hype and hypocrisy of generative artificial intelligence ({AI}) rhetoric,'' \emph{M/C Journal}, vol.~26, 10 2023.

\bibitem{mohsenimofidi2025context}
S.~Mohsenimofidi, M.~Galster, C.~Treude, and S.~Baltes, ``Context engineering for {AI} agents in open-source software,'' \emph{arXiv preprint arXiv:2510.21413}, 2025.

\bibitem{digitalanthro}
H.~A. Horst and D.~Miller, \emph{Digital anthropology}.\hskip 1em plus 0.5em minus 0.4em\relax Routledge, 2020.

\bibitem{mcluhan1977laws}
M.~McLuhan, ``Laws of the media,'' \emph{ETC: A Review of General Semantics}, pp. 173--179, 1977.

\bibitem{hou2023large}
X.~Hou, Y.~Zhao, Y.~Liu, Z.~Yang, K.~Wang, L.~Li, X.~Luo, D.~Lo, J.~Grundy, and H.~Wang, ``Large language models for software engineering: A systematic literature review,'' \emph{ACM Transactions on Software Engineering and Methodology}, vol.~33, no.~8, pp. 1--79, 2024.

\bibitem{xu2023we}
B.~Xu, T.-D. Nguyen, T.~Le-Cong, T.~Hoang, J.~Liu, K.~Kim, C.~Gong, C.~Niu, C.~Wang, B.~Le \emph{et~al.}, ``Are we ready to embrace generative {AI} for software {Q\&A}?'' in \emph{Proceedings of the International Conference on Automated Software Engineering}.\hskip 1em plus 0.5em minus 0.4em\relax IEEE, 2023, pp. 1713--1717.

\bibitem{shawar2007chatbots}
B.~A. Shawar and E.~Atwell, ``Chatbots: Are they really useful?'' \emph{Journal for Language Technology and Computational Linguistics}, vol.~22, no.~1, pp. 29--49, 2007.

\bibitem{hicks2024new}
C.~M. Hicks, C.~Lee, and K.~Foster-Marks, ``The new developer: {AI} skill threat, identity change \& developer thriving in the transition to {AI}-assisted software development,'' 2024.

\bibitem{russo2024generative}
D.~Russo, S.~Baltes, N.~van Berkel, P.~Avgeriou, F.~Calefato, B.~Cabrero-Daniel, G.~Catolino, J.~Cito, N.~Ernst, T.~Fritz \emph{et~al.}, ``Generative {AI} in software engineering must be human-centered: The {Copenhagen} {Manifesto},'' \emph{Journal of Systems and Software}, vol. 216, p. 112115, 2024.

\bibitem{latour2007reassembling}
B.~Latour, \emph{Reassembling the social: An introduction to actor-network-theory}.\hskip 1em plus 0.5em minus 0.4em\relax Oup Oxford, 2007.

\bibitem{ferino2025novice}
S.~Ferino, R.~Hoda, J.~Grundy, and C.~Treude, ``Novice developers’ perspectives on adopting {LLMs} for software development: A systematic literature review,'' \emph{arXiv preprint arXiv:2503.07556}, 2025.

\bibitem{mustafa}
M.~Suleyman, \emph{\BIBforeignlanguage{en}{The coming wave}}.\hskip 1em plus 0.5em minus 0.4em\relax New York, NY: Crown Publishing Group, Sep. 2023.

\bibitem{xiao2025self}
T.~Xiao, Y.~Fan, F.~Calefato, C.~Treude, R.~G. Kula, H.~Hata, and S.~Baltes, ``Self-admitted {GenAI} usage in open-source software,'' \emph{arXiv preprint arXiv:2507.10422}, 2025.

\bibitem{butler2025dear}
J.~Butler, J.~Suh, S.~Haniyur, and C.~Hadley, ``Dear diary: A randomized controlled trial of generative {AI} coding tools in the workplace,'' in \emph{Proceedings of the International Conference on Software Engineering: Software Engineering in Practice}.\hskip 1em plus 0.5em minus 0.4em\relax IEEE, 2025, pp. 319--329.

\bibitem{terragni2025future}
V.~Terragni, A.~Vella, P.~Roop, and K.~Blincoe, ``The future of {AI}-driven software engineering,'' \emph{ACM Transactions on Software Engineering and Methodology}, vol.~34, no.~5, pp. 1--20, 2025.

\bibitem{pimenova2025good}
V.~Pimenova, S.~Fakhoury, C.~Bird, M.-A. Storey, and M.~Endres, ``Good vibrations? {A} qualitative study of co-creation, communication, flow, and trust in vibe coding,'' \emph{arXiv preprint arXiv:2509.12491}, 2025.

\bibitem{treude2025bot}
C.~Treude and C.~M. Poskitt, ``Bot-driven development: From simple automation to autonomous software development bots,'' in \emph{Proceedings of the International Workshop on Bots in Software Engineering}, 2025, pp. 18--22.

\bibitem{he2025llm}
J.~He, C.~Treude, and D.~Lo, ``{LLM}-based multi-agent systems for software engineering: Literature review, vision, and the road ahead,'' \emph{ACM Transactions on Software Engineering and Methodology}, vol.~34, no.~5, pp. 1--30, 2025.

\bibitem{Naur}
P.~Naur, ``Programming as theory building,'' \emph{Microprocessing and Microprogramming}, vol.~15, no.~5, pp. 253--261, 1985.

\bibitem{treude2025interacting}
C.~Treude and R.~G. Kula, ``Interacting with {AI} reasoning models: Harnessing "thoughts" for {AI}-driven software engineering,'' \emph{arXiv preprint arXiv:2503.00483}, 2025.

\bibitem{programmingbyexample}
D.~C. Halbert, \emph{Programming by example}.\hskip 1em plus 0.5em minus 0.4em\relax University of California, Berkeley, 1984.

\bibitem{VRsoftware}
J.~M. Gonzalez-Barahona, ``{IDEs} in the age of {LLMs} and {XR},'' in \emph{Proceedings of the Workshop on Integrated Development Environments}, 2024, pp. 66--69.

\bibitem{baltes2025guidelines}
S.~Baltes, F.~Angermeir, C.~Arora, M.~M. Bar{\'o}n, C.~Chen, L.~B{\"o}hme, F.~Calefato, N.~Ernst, D.~Falessi, B.~Fitzgerald \emph{et~al.}, ``Guidelines for empirical studies in software engineering involving large language models,'' \emph{arXiv preprint arXiv:2508.15503}, 2025.

\bibitem{poth}
C.~N. Poth, \emph{Innovation in mixed methods research: A practical guide to integrative thinking with complexity}.\hskip 1em plus 0.5em minus 0.4em\relax Los Angeles: SAGE, 2018.

\bibitem{hicks}
C.~M. Hicks, C.~S. Lee, and M.~Ramsey, ``Developer thriving: Four sociocognitive factors that create resilient productivity on software teams,'' \emph{IEEE Software}, vol.~41, no.~4, pp. 68--77, 2024.

\bibitem{shallows}
N.~Carr, \emph{The shallows: What the {Internet} is doing to our brains}.\hskip 1em plus 0.5em minus 0.4em\relax WW Norton \& Company, 2020.

\bibitem{ahmed2025can}
T.~Ahmed, P.~Devanbu, C.~Treude, and M.~Pradel, ``Can {LLMs} replace manual annotation of software engineering artifacts?'' in \emph{Proceedings of the International Conference on Mining Software Repositories}, 2025, pp. 526--538.

\bibitem{bano2024large}
M.~Bano, R.~Hoda, D.~Zowghi, and C.~Treude, ``Large language models for qualitative research in software engineering: Exploring opportunities and challenges,'' \emph{Automated Software Engineering}, vol.~31, no.~1, p.~8, 2024.

\bibitem{steinmacher}
M.~Gerosa, B.~Trinkenreich, I.~Steinmacher, and A.~Sarma, ``Can {AI} serve as a substitute for human subjects in software engineering research?'' \emph{Automated Software Engineering}, vol.~31, no.~1, p.~13, 2024.

\bibitem{harrison1996editorial}
W.~Harrison and V.~R. Basili, ``Editorial,'' \emph{Empirical Software Engineering}, vol.~1, no.~1, pp. 5--10, 1996.

\bibitem{abou2022software}
Z.~Abou~Khalil and S.~Zacchiroli, ``Software artifact mining in software engineering conferences: A meta-analysis,'' in \emph{Proceedings of the International Symposium on Empirical Software Engineering and Measurement}, 2022, pp. 227--237.

\bibitem{shi2022buglistener}
L.~Shi, F.~Mu, Y.~Zhang, Y.~Yang, J.~Chen, X.~Chen, H.~Jiang, Z.~Jiang, and Q.~Wang, ``Buglistener: {I}dentifying and synthesizing bug reports from collaborative live chats,'' in \emph{Proceedings of the International Conference on Software Engineering}, 2022, pp. 299--311.

\bibitem{lu2023llama}
J.~Lu, L.~Yu, X.~Li, L.~Yang, and C.~Zuo, ``{LLaMA-Reviewer}: Advancing code review automation with large language models through parameter-efficient fine-tuning,'' in \emph{Proceedings of the International Symposium on Software Reliability Engineering}.\hskip 1em plus 0.5em minus 0.4em\relax IEEE, 2023, pp. 647--658.

\bibitem{zhang2024automatic}
Y.~Zhang, Z.~Qiu, K.-J. Stol, W.~Zhu, J.~Zhu, Y.~Tian, and H.~Liu, ``Automatic commit message generation: A critical review and directions for future work,'' \emph{IEEE Transactions on Software Engineering}, pp. 816--835, 2024.

\bibitem{zhong2024can}
L.~Zhong and Z.~Wang, ``Can {LLM} replace {Stack Overflow}? {A} study on robustness and reliability of large language model code generation,'' in \emph{Proceedings of the AAAI Conference on Artificial Intelligence}, vol.~38, no.~19, 2024, pp. 21\,841--21\,849.

\bibitem{cambiaso2023scamming}
E.~Cambiaso and L.~Caviglione, ``Scamming the scammers: Using {ChatGPT} to reply mails for wasting time and resources,'' \emph{arXiv preprint arXiv:2303.13521}, 2023.

\bibitem{liu2024your}
J.~Liu, C.~S. Xia, Y.~Wang, and L.~Zhang, ``Is your code generated by {ChatGPT} really correct? {Rigorous} evaluation of large language models for code generation,'' \emph{Advances in Neural Information Processing Systems}, vol.~36, 2024.

\bibitem{xue2024domain}
Z.~Xue, L.~Li, S.~Tian, X.~Chen, P.~Li, L.~Chen, T.~Jiang, and M.~Zhang, ``Domain knowledge is all you need: A field deployment of {LLM}-powered test case generation in {FinTech} domain,'' in \emph{Proceedings of the International Conference on Software Engineering: Companion Proceedings}, 2024, pp. 314--315.

\bibitem{conrardy2024image}
A.~Conrardy and J.~Cabot, ``From image to {UML}: First results of image based {UML} diagram generation using {LLMs},'' \emph{arXiv preprint arXiv:2404.11376}, 2024.

\bibitem{10.1145/3762183}
H.~Y. Lin, P.~Thongtanunam, C.~Treude, M.~W. Godfrey, C.~Liu, and W.~Charoenwet, ``Leveraging reviewer experience in code review comment generation,'' \emph{ACM Transactions on Software Engineering and Methodology}, Aug. 2025.

\bibitem{nguyen2024encoding}
H.~Nguyen, P.~Thongtanunam, and C.~Treude, ``Encoding version history context for better code representation,'' in \emph{Proceedings of the International Conference on Mining Software Repositories}, 2024, pp. 631--636.

\bibitem{treude2023she}
C.~Treude and H.~Hata, ``She elicits requirements and he tests: Software engineering gender bias in large language models,'' in \emph{Proceedings of the International Conference on Mining Software Repositories}.\hskip 1em plus 0.5em minus 0.4em\relax IEEE, 2023, pp. 624--629.

\bibitem{sami2023case}
M.~Sami, A.~Sami, and P.~Barclay, ``A case study of fairness in generated images of large language models for software engineering tasks,'' in \emph{Proceedings of the International Conference on Software Maintenance and Evolution}.\hskip 1em plus 0.5em minus 0.4em\relax IEEE, 2023, pp. 391--396.

\bibitem{davies2014s}
S.~Davies and M.~Roper, ``What's in a bug report?'' in \emph{Proceedings of the International Symposium on Empirical Software Engineering and Measurement}, 2014, pp. 1--10.

\bibitem{alali2008s}
A.~Alali, H.~Kagdi, and J.~I. Maletic, ``What's a typical commit? {A} characterization of open source software repositories,'' in \emph{Proceedings of the International Conference on Program Comprehension}.\hskip 1em plus 0.5em minus 0.4em\relax IEEE, 2008, pp. 182--191.

\bibitem{li2017they}
Z.-X. Li, Y.~Yu, G.~Yin, T.~Wang, and H.-M. Wang, ``What are they talking about? {Analyzing} code reviews in pull-based development model,'' \emph{Journal of Computer Science and Technology}, vol.~32, pp. 1060--1075, 2017.

\bibitem{treude2025how}
C.~Treude and M.~A. Gerosa, ``How developers interact with {AI}: A taxonomy of human-{AI} collaboration in software engineering,'' in \emph{Proceedings of the International Conference on AI Foundation Models and Software Engineering}, 2025, pp. 236--240.

\bibitem{xiao2024devgpt}
T.~Xiao, C.~Treude, H.~Hata, and K.~Matsumoto, ``{DevGPT}: Studying developer-{ChatGPT} conversations,'' in \emph{Proceedings of the International Conference on Mining Software Repositories}, 2024, pp. 227--230.

\bibitem{liu2024no}
Z.~Liu, Y.~Tang, X.~Luo, Y.~Zhou, and L.~F. Zhang, ``No need to lift a finger anymore? {Assessing} the quality of code generation by {ChatGPT},'' \emph{IEEE Transactions on Software Engineering}, vol.~50, pp. 1548--1584, 2024.

\bibitem{yetistiren2022assessing}
B.~Yetistiren, I.~Ozsoy, and E.~Tuzun, ``Assessing the quality of {GitHub} {Copilot’s} code generation,'' in \emph{Proceedings of the International Conference on Predictive Models and Data Analytics in Software Engineering}, 2022, pp. 62--71.

\bibitem{parnin2012crowd}
C.~Parnin, C.~Treude, L.~Grammel, and M.-A. Storey, ``Crowd documentation: Exploring the coverage and the dynamics of {API} discussions on {Stack Overflow},'' \emph{Georgia Institute of Technology, Technical Report}, vol.~11, 2012.

\bibitem{mixedthreats}
X.~Yu and D.~Khazanchi, ``Using embedded mixed methods in studying is phenomena: Risks and practical remedies with an illustration,'' \emph{Communications of the Association for Information Systems}, vol.~41, p.~2, 2017.

\end{thebibliography}
\end{document}